\documentclass[pre,twocolumn,twoside,floatfix,longbibliography]{revtex4-2}
\usepackage{amsmath}
\usepackage{graphicx}
\usepackage{hyperref}
\hypersetup{
	unicode=false,                                                                 
	pdffitwindow=false,                                                            
	pdfstartview={FitH},                                                           
	pdftitle={Memory-cognizant generalization to Simon's random-copying neutral model}, 
	pdfauthor={Joseph D.~O'Brien and James P.~Gleeson},                
	pdfsubject={},                                                                 
	pdfnewwindow=true,                                                             
	colorlinks=true,                                                               
	linkcolor=blue,                                                                
	citecolor=blue,                                                                
	filecolor=magenta,                                                             
	urlcolor=blue                                                                  
}
\usepackage{booktabs, threeparttable, cellspace, tabularx}
\usepackage{varwidth}
\newcommand{\btext}[1]{\text{\begin{varwidth}[t]{\textwidth}\centering#1\end{varwidth}}}
\usepackage{bbm}
\usepackage{hhline}
\graphicspath{ {Images/} }
\usepackage{caption}
\usepackage{tikz}
\usetikzlibrary{calc,decorations.markings}
\newcommand*{\pd}[3][]{\ensuremath{\frac{\partial^{#1} #2}{\partial #3}}}

\begin{document}
	
	\title{Memory-cognizant generalization to Simon's random-copying neutral model}
	\author{Joseph D.~\surname{O'Brien}}
	\affiliation{MACSI, Department of Mathematics and Statistics, University of Limerick, Limerick V94 T9PX, Ireland}
	\author{James P.~\surname{Gleeson}}
	\affiliation{MACSI, Department of Mathematics and Statistics, University of Limerick, Limerick V94 T9PX, Ireland}

	\begin{abstract}
		Simon's classical random-copying model, introduced in 1955, has garnered much attention for its ability, in spite of an apparent simplicity, to produce characteristics similar to those observed across the spectrum of complex systems. Through a discrete-time mechanism in which items are added to a sequence based upon rich-gets-richer dynamics, Simon demonstrated that the resulting size distributions of such sequences exhibit  power-law tails. The simplicity of this model arises from the approach by which copying occurs uniformly over all previous elements in the sequence. Here we propose a generalization of this model which moves away from this uniform assumption,  instead incorporating memory effects that allow the copying event to occur via an arbitrary age-dependent kernel. Through this approach we first demonstrate the potential to determine further information regarding the structure of sequences from the classical model  before illustrating, via analytical study and numeric simulation, the flexibility offered by the arbitrary choice of memory. Furthermore we demonstrate how previously proposed memory-dependent models can be further studied as specific cases of the proposed framework.
	\end{abstract}
	\maketitle
	\section{Introduction}
	
	Within natural systems from an assortment of domains there are underlying properties which are found to consistently appear. These occurrences result in researchers aiming to develop general theories which can capture aspects of these phenomena independent of the domain under consideration~\cite{Gisiger2001,west2017scale}. One such instance is  the distribution underlying the abundance of a system's constituents, which is frequently described by a highly right-skewed distribution~\cite{Willis1922,Simon1955,Newman2005}. This apparently ubiquitous phenomena has been observed in a variety of domains,  including baby-name popularity~\cite{Hahn2003, Bentley2004}, citations to scientific literature~\cite{DeSollaPrice1965, Redner2005,Simkin_2007, Radicchi2008}, user-generated passwords~\cite{Wang2017}, and the market share of different cryptocurrencies~\cite{Elbahrawy2017}.
	
	Motivated by this apparently ubiquitous property, researchers have investigated  mechanisms which may reproduce such heavy-tailed size distributions. These models tend to create a population of elements, each of which has a certain identity or variant, who reproduce over time through a given process, the abundances of said variants are generally the quantities of interest. Typically these mechanisms also offer the possibility of mutation or the creation of a new type of variant in a given reproduction event. The most common of these frameworks are known as \textit{neutral models} due to the assumption that each variant in the population has equal fitness~\cite{Azaele2016,Martinello2017}.  This neutrality implies that there are no intrinsic advantage offered to any type of variant within the model. For a detailed overview of the most common of these models, alongside their limitations, we refer the interested reader to Ref.~\cite{Leroi2020}.
	
	In 1955 Herbert Simon, building upon the earlier work of G.~U.~Yule~\cite{Yule1925}, famously introduced such a neutral model based upon random-copying (reproduction was viewed as a copying event) with mutation (which he referred to as an innovation event) framework that could reproduce the power-law distributions observed within empirical systems~\cite{Simon1955}. The fundamental property of this model is that the likelihood of the next element being of a certain variant is dependent upon the number of previous occurrences of this variant. Notwithstanding the model's apparent simplicity, it has been shown to accurately describe the distribution of abundances within a number of complex systems, including the citation dynamics of scientific literature~\cite{Newman2009, Dodds2017}, family-names~\cite{Zanette2001}, and the growth of both the world wide web~\cite{Bornholdt2001} and open-source software developments~\cite{Maillart2008}.
	
	Similar models have appeared throughout the literature since Simon's original article --- in particular those placing emphasis upon the cumulative advantage or rich-get-richer mechanisms. Arguably the most famous such approach has been in the development of network growth models within the field of complex networks initiated by Barab\'asi and Albert in 1999~\cite{Barabasi1999} (although the first such framework was produced through Price's model of citation growth~\cite{DeSollaPrice1965, Price1976}). In spite of the misleading use of the word `advantage' within these frameworks it is important to highlight, as mentioned in~\cite{Leroi2020}, that these models are in fact neutral models as the nodes have no inherent fitness: their likelihood of selection is determined only by the number of times they have previously been selected, although there have been a number of extensions proposed that move beyond from this neutral assumption~\cite{Bianconi2001,Caldarelli2002}.
	
	As demonstrated through the preceding commentary, Simon's model has proven extremely adaptable in describing behaviors within a wide spectrum of  domains. Remarkably, in spite of the model's supposed straightforwardness, the research community is still providing new insights into its behaviors. For example, analytical quantification of the first-mover advantage offered towards the initial variant in Simon's model was demonstrated in~\cite{Dodds2017}, while the effect of averaging across multiple realizations of a closely related model upon the distribution of larger abundances, denoted as peloton dynamics, have also been considered~\cite{Hebert-Dufresne2012}. There have also been a number of studies which have incorporated memory-effects within Simon's model such that the likelihood of a variant being copied is no longer uniform across all previous elements but rather is dependent upon how long ago the variant was last used. Such a framework requires the incorporation of a probability function~$\phi(\tau, t)$ describing the likelihood of copying at time~$t$ an element which appeared at time~$\tau$. For example, in Refs.~\cite{Cattuto2006, Cattuto2007} the authors introduce a fat-tailed memory distribution to the model which could capture long-range correlations in times between copying events. Furthermore, the authors in Ref.~\cite{Schaigorodsky2018} consider a finite-sized memory kernel such that if a certain time has passed since a variant has last been copied it becomes extinct. Reference~\cite{Bertoin_2019} studies the scenario whereby the likelihood of copying is dependent on a power function of its abundance, with the asymptotic behavior of said abundance being analyzed via a branching process interpretation. Lastly, a specific age-dependent model described by an additional parameter is again studied in terms of its branching behavior in Ref.~\cite{baur2020twoparameter}. Questions remain however as to the effect of incorporating a more general memory function, depending on the time since an element appeared, to Simon's model. 
	
	With these questions in mind, in this article we consider a branching process approach~\cite{Harris1963,Athreya2013} towards describing a generalized Simon's model. Such techniques have previously proven extremely useful in representing similar random-copying phenomena in online diffusion scenarios~\cite{van_der_Lans_2010,Iribarren2011,Yagan_2013,Gleeson2014, Gleeson_2016, OBrien_2019}. Initially, we take this approach in describing the classical Simon's model and demonstrate how the interpretation allows numerous quantities describing the distribution of abundances within the process to be obtained in an analytically tractable manner. Moreover, we obtain results regarding the temporal evolution of a given variant as a consequence of the model incorporating the time at which a given element first appears. We proceed to use a similar framework to study a general memory function which depends upon the time $t-\tau$ elapsed since an element appeared to produce a generalized Simon's model (GSM) and demonstrate how statistical properties of abundances arising through such a process may be obtained for an arbitrary choice of said memory function. We also highlight how the previous literature focusing on specific choices of memory function~\cite{Cattuto2006,Cattuto2007,Schaigorodsky2018}, which are viewed as particular scenarios of our proposed framework, may be further understood using the branching process interpretation proposed here. 
	
	The remainder of the paper is laid out as follows. In Section~\ref{sec:classic} we introduce Simon's original model in detail before demonstrating a branching process approach towards describing it from which classic, alongside new, results are derived. Having demonstrated the usefulness of our framework we proceed in Section~\ref{sec:GSM} to introduce a generalization of Simon's model  incorporating arbitrary age-dependent memory effects and present a thorough analysis of its features. In Section~\ref{sec:simulations} we proceed to validate the GSM framework through extensive numerical simulations. We demonstrate how previous extensions to Simon's model may be viewed further studied using the framework presented here resulting in additional information regarding said extensions now being attainable in Section~\ref{sec:special} before drawing our conclusions in Section~\ref{sec:conclusions}.
	
	\section{Results from Simon's classical model - new and old}\label{sec:classic}
	
	The model devised by Simon~\cite{Simon1955}, like most neutral models, is remarkable for both its apparent simplicity and the  variety of behaviors it exhibits. 
	The original representation describes an author creating a body of text that is the population of interest, with every word used representing an element:each unique word corresponds to a  variant in the neutral model framework. In the analysis to follow we shall use the neutral model descriptors (e.g., population, element, variant). The system initializes at time~$t = 1$ with a single element and then at each subsequent discrete time-step of length~$\Delta t$ another element is added. How the variant of this element is chosen occurs through a probabilistic framework whereby there may be a mutation (innovation event in the original model) with probability~(w.p.)~$\mu$ such that this new element is a new variant in the population. Alternatively, occurring w.p.~$1-\mu$, the new element is from a reproduction event implying a random replication from one of the previous elements. Our main focus in  this article is on considering how the new element chooses a previous element to copy from; initially  however, we study the case considered by Simon whereby this choice is made uniformly over all previously used elements. 	
	
	\begin{figure}
		\centering
		\includegraphics[width = \columnwidth]{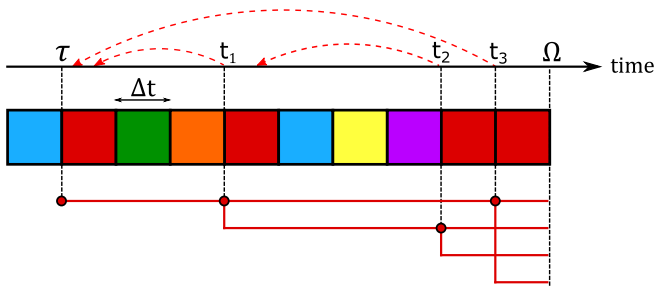}
		\caption{\textbf{Schematic illustration of the classical Simon's model and its corresponding branching structure.} At each discrete time interval of length~$\Delta t$ a new element is created either via reproduction such that it copies the variant of a previous element or mutation where it is an entirely novel variant. The variants are represented here by colors, and we focus on the dynamics of the red variant that first appeared via mutation at time~$\tau$. A reproduction event occurs at time $t_1$ whereby the new element randomly chooses from all those which had appeared previously (indicated by the dashed line) to determine its type of variant, which is the red element of interest. This results in a new branch being created in the branching process interpretation (bottom plot). The branch may  itself  reproduce in the future, as occurs at time $t_2$ while the original element may also still produce progeny, as occurs at time $t_3$. The quantity of interest is the abundance of this variant at time $\Omega$: in the realization shown, the final abundance (total number of red elements in the sequence) is four.}
		\label{fig:branching_simon}
	\end{figure}
	
	We now focus on the main quantity of interest which is the number of times~$n$ that a variant which first appeared at time~$\tau$ has been used by time~$\Omega$. Of course, due to the stochastic nature of the model, this quantity is a random variable and so we define $q_n(\tau, \Omega)$ to describe the corresponding probability of it taking the value $n$. More specifically, we shall interact with this quantity through its corresponding probability generating function (PGF)~\cite{Wilf2005} given by
	\begin{equation}
		H(\tau, \Omega; x) = \sum_{n=1}^{\infty} q_n(\tau, \Omega) \, x^n,
	\end{equation}
	with final condition $H(\Omega,\Omega;x) = x$, i.e., a variant which is created through a mutation event just as the process concludes would have an abundance of one. We now examine the functional form of the PGF that results from the mechanisms of Simon's model (see Fig.~\ref{fig:branching_simon}). In particular, we consider how the distribution of a variant's abundance may change over a short time interval~$(\tau-\Delta t, \tau)$. Within this window there are three possible options from which the variant of a new element may be determined:
	
	\begin{enumerate}
		\item First, the new element may be a result of a reproduction event w.p.~$1-\mu$ and the element which we are focusing on is chosen w.p.~$\Delta t/\tau$, this being the time during which the element was present $\Delta t$ divided by the entire time $\tau$ of the process until this point. The reproduction event results in a further element of the variant which itself may reproduce in the future thus contributing~$\left[H(\tau, \Omega; x)\right]^2$ to the PGF, where we have used the fact that both elements are now identical due to the variant abundance being the only factor influencing future copying events alongside the property that the PGF of the sum of two random variables is simply the product of their corresponding PGFs~\cite{Wilf2005}.
		\item Second, the new element may again be a result of reproduction w.p.~$1 - \mu$ but with the source of the reproduction (copying) being another element rather than the one under consideration. Such an copying event occurs w.p. $(\tau - \Delta t)/\tau$, i.e., the fraction of time the other elements were present for. In this scenario a further branch is not created and as such the contribution to the PGF of interest is simply $H(\tau, \Omega; x)$, i.e., a single branch that may result in future copying events.
		\item Finally, the new element may be as a result of a mutation, w.p. $\mu$. This of course results in a different variant to that under consideration and as such again contributes $H(\tau, \Omega; x)$ to the PGF.  
	\end{enumerate}
	
	\begin{widetext}
		Taking these terms and their corresponding probabilities together we produce the following difference equation describing the time evolution of the PGF
		\begin{equation}
			H(\tau - \Delta t, \Omega; x) = (1 - \mu)\left(\frac{\Delta t}{\tau}\right)\left[H(\tau, \Omega ;x)\right]^2 +
			(1 - \mu)\left(\frac{\tau - \Delta t}{\tau}\right)H(\tau, \Omega ;x) + \mu H(\tau, \Omega; x), 
			\label{clas_time_change}
		\end{equation} and if we proceed to consider the scenario in which $\Delta t \ll \tau$ we may move to a continuum time limit that proves  more tractable for mathematical analysis:
		\begin{equation}
			-\frac{d H}{d \tau} = \frac{1 - \mu}{\tau}\left(H^2 - H\right).
			\label{clas_ode}
		\end{equation}
	\end{widetext}
	This is an ordinary differential equation under the assumption of fixed $\Omega$, where $H = H(\tau,\Omega;x)$. This equation is in fact of the Ricatti form which, when combined with the aforementioned final condition, may be solved to obtain an exact representation of the PGF's functional form by
	\begin{equation}
		H(\tau, \Omega; x) = \frac{x}{x\left[ 1 - (\frac{\tau}{\Omega})^{-(1-\mu)} \right] + (\frac{\tau}{\Omega})^{-(1-\mu)}}.
		\label{eq:Hclassic}
	\end{equation}
	With this solution for the PGF at hand we may proceed to conduct analysis regarding the underlying distribution of variant abundance.
	
	\subsection{Analytically obtained statistical properties}
	
	The advantage of having obtained a functional representation regarding the dynamics of Simon's model is that it provides information regarding the entire distribution of abundances and is also extremely amenable to quantitative analysis. We now proceed to highlight some, although far from all, of these possible analyses regarding the statistical behavior of the variants' abundance.
	
	\subsubsection{Distribution of variant abundance for given seed time}
	
	First of all we consider the probability distribution of a variant's abundance for a given seed-time. Specifically we focus on the previously defined probability $q_n(\tau, \Omega)$ which may be accessed through the PGF $H$ via the well-known property
	\begin{align}
		q_n(\tau, \Omega) = \left.\frac{1}{(n-1)!}\pd[n]{}{x^n}\left[\frac{H(\tau,\Omega;x)}{x}\right]\right\rvert_{x = 0},
		\label{eq:qn_pgf}
	\end{align} 
	such that knowledge of the $n$-th derivative of Eq.~\eqref{eq:Hclassic} is required. Generally this distribution is completed numerically or, to avoid the inaccuracies caused from numerical differentiation, via transformation to the complex plane from which Cauchy's theorem may be used~\cite{Abate1992, Gleeson2014}. Fortunately, in this specific case, it can be seen that the PGF in fact describes a geometric distribution (with an additional power of $x$) such that the probability mass function can be written explicitly as
	\begin{equation}
		q_n(\tau,\Omega) = \left(\frac{\tau}{\Omega}\right)^{1-\mu}\left[1 - \left(\frac{\tau}{\Omega}\right)^{1-\mu}\right]^{n-1}.
		\label{eq:qn_classic}
	\end{equation} 
	An array of analysis is possible with the analytical distribution at hand, for example we may consider the likelihood that a variant which first appeared at time $\tau$ does not appear again by time $\Omega$ which is given by
	\begin{equation}
		q_1(\tau,\Omega) = \left(\frac{\tau}{\Omega}\right)^{1-\mu}.
		\label{eq:q1_classic}
	\end{equation} 
	Furthermore, we may readily calculate the moments of abundance size directly from the distribution, however, with the analysis to come later in this article in mind, we will now focus on obtaining such quantities directly from the PGF.
	
	\subsubsection{Moments of variant abundance distribution}\label{subsubsec:classic_mean}
	
	We first consider the average number of times~$m(\tau, \Omega)$ that a variant that first appeared at time~$\tau$ has appeared across all elements by time~$\Omega$. Taking advantage of the fact that the  distribution's moments are readily attainable through repeated differentiation of Eq.~\eqref{eq:Hclassic} w.r.t.~$x$, the average or first moment is obtained through the following calculation
	\begin{align}
		m(\tau, \Omega) &= \left. \pd{H(\tau, \Omega; x)}{x} \right\rvert_{x = 1} \nonumber \\
		&= \left(\frac{\tau}{\Omega}\right)^{-(1-\mu)}.
		\label{eq:mean_classic}
	\end{align}
	Interestingly, we note that this quantity is the inverse of the probability that the variant had only appeared once in the population given by Eq.~\eqref{eq:q1_classic}.
	This expression highlights a number of facets within Simon's model,  most obviously the `early-mover' advantage  suggesting that variants with earlier seed times $\tau$ will have, on average, appeared more times in a given realization of the process. We specifically highlight that the mean abundance of a variant in fact exhibits a power-law relationship with the seed time. 
	
	One final point we comment on regarding this moment is the average number of appearances for the variant describing the introductory element seeded at time~$\tau = 1$, which is given by $(1/\Omega)^{-(1-\mu)}$. This expression is in general agreement with the analysis found in~\cite{Dodds2017} (assuming that~$\mu~\ll~1$ such that~$1/\Gamma(2-\mu) \approx 1$) demonstrating the intrinsic advantage offered to the first variant to appear in a realization of Simon's model. 
	
	Of course we may obtain further moments of the distribution through similar calculations to those shown above. For example, the variance of variant abundance $v(\tau, \Omega)$ may be found by observing that
	\begin{align}
		v(\tau, \Omega) &=  \left. \pd[2]{H(\tau, \Omega; x)}{x^2}  +  \pd{H(\tau, \Omega; x)}{x} - \left[\pd{H(\tau, \Omega; x)}{x}\right]^2\right\rvert_{x = 1} \nonumber \\
		&= \left(\frac{\tau}{\Omega}\right)^{-(1-\mu)}\left[\left(\frac{\tau}{\Omega}\right)^{-(1-\mu)} - 1\right],
	\end{align}
	which allows analysis regarding the dispersal of the abundance size to be readily calculated.
	
	\subsubsection{Distribution of variant abundances across all seed times}
	Simon's model originally found fame for its ability to generate power-law distributed popularity values when all variants within a given sequence are considered, regardless of their ages. This distribution $\tilde{q}_n(\Omega)$ thus focuses on all seed times rather than the distribution of variants with a given seed time as our preceding analysis has considered. The classic result is however directly obtainable from our  analysis by taking the distribution for a given seed time from Eq.~\eqref{eq:qn_classic} and averaging over all seed times, which are uniformly distributed over the interval $[0,\Omega]$, to obtain
	\begin{align}
		\tilde{q}_n &= \frac{1}{\Omega}\int_{0}^{\Omega} q_n(\tau,\Omega) \, \mathrm{d}\tau \nonumber \\
		&=  \frac{1}{\Omega}\int_{0}^{\Omega} \left(\frac{\tau}{\Omega}\right)^{1-\mu}\left[1 - \left(\frac{\tau}{\Omega}\right)^{1-\mu}\right]^{n-1} \, \mathrm{d}\tau \nonumber \\
		&\sim \frac{1}{1 - \mu} \, \beta\left[\frac{2-\mu}{1-\mu}, n \right], \qquad \text{for large $n$}
	\end{align} 
	where~$\beta$ represent the beta distribution and by consequently making use of the approximation~$\beta(a,b) \sim b^{-a}$ for large~$b$, we obtain
	\begin{equation}
		\tilde{q}_n \sim \frac{1}{1 - \mu} n^{-\left(\frac{2-\mu}{1-\mu}\right)},
		\label{eq:large_qn_classic}
	\end{equation}	
	the power-law distribution with exponent larger than two governing the entire population sequence as originally demonstrated by Simon.
	
	\subsection{Numerical simulations}\label{subsec:classic_sim}
	
	We now provide results from numerical simulations of the classical Simon's model with the aim of validating the branching process approach described in the preceding section. Specifically, we compute an ensemble of $10^6$ realizations of the process with length~$\Omega$ before considering the quantities derived above. In general, when considering a given seed time~$\tau$ we force a variant to be created at this time but otherwise the simulations are exactly as described by Simon in his original work.
	\begin{figure*}
		\centering
		\includegraphics[width = \textwidth]{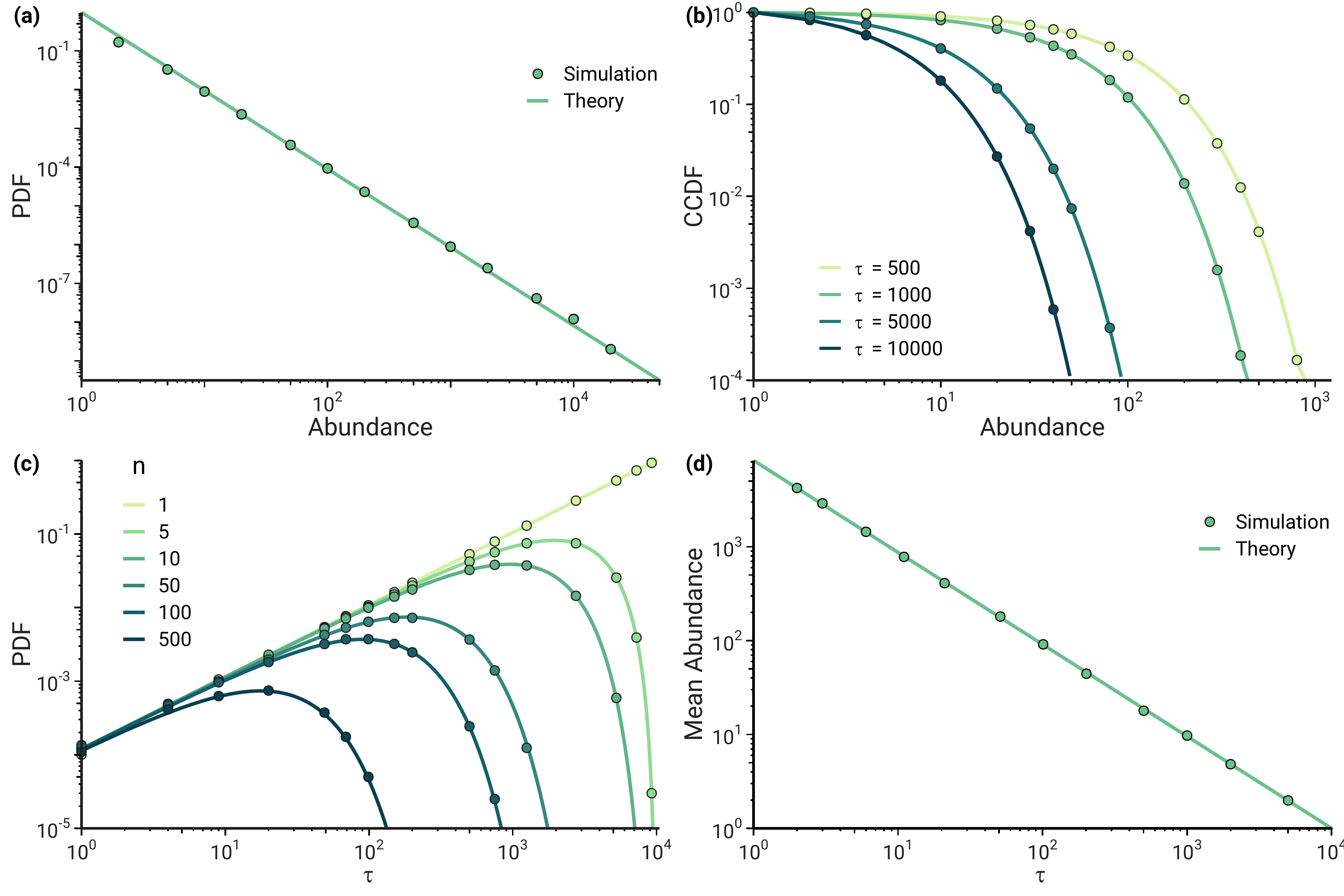}
		\caption{\textbf{Simulations and corresponding analytical results for the classical Simon's model. (a)~}Probability distribution across all abundances of variants within realizations of Simon's model with~$\Omega = 5 \times 10^5$ and~$\mu = 0.01$. \textbf{(b)}~The complementary cumulative distribution of variant abundance based upon their seed time ($\tau = 5\times10^2, 10^3, 5\times10^3,$ and $10^4$) where in each case the observation time is $2.5\times10^4$ time units after the seed time. The early-mover advantage is observed: those variants with earlier seed times have larger likelihood of larger abundance despite all having the same age. \textbf{(c)}~Probability distribution function of variant abundance dependent upon the variant's seed time for a number of different abundance values with $\Omega = 10^5$. \textbf{(d)}~The average popularity of a variant with $\Omega = 10^5$ in each case for multiple seed times is shown. The power-law relationship with seed time is demonstrated, alongside theoretical values given by Eq.~\eqref{eq:mean_classic}. Each of the above simulations are performed over $10^6$ ensembles.}
		\label{fig:class_simon_sims}
	\end{figure*}	
	We first consider the overall distribution of variant abundance originally shown to result in a power-law distribution for larger abundances and as described  through Eq.~\eqref{eq:large_qn_classic}. The corresponding distribution is shown in Fig.~\ref{fig:class_simon_sims}(a) where simulations have been performed with $\Omega = 5 \times 10^5$ and $\mu = 0.01$.
	
	We now consider the first original result obtained as a consequence of our approach. Specifically, the distribution of abundances for a variant seeded at time $\tau$ given by Eq.~\eqref{eq:qn_classic} is considered for a number of different seed times $\tau = 5\times10^2, 10^3, 5\times10^3,$ and $10^4$. We then observe the distribution of abundances observed after $2.5\times10^4$ time units in each scenario such that each seed time is considered with the same age at observation time $\Omega = \tau + 2.5\times10^4$. The results of said simulations are shown in Fig.~\ref{fig:class_simon_sims}(b) where we observe the `early mover' advantage offered by Simon's model which has been a motivating factor in similar preferential attachment based models. Figure~\ref{fig:class_simon_sims}(c) considers a similar framework but instead focuses on how the probability of a variant having abundance $n$ varies as a function of seed time, where in each case $\Omega = 10^5$. We again observe the early-mover advantage here with those large $n$ values being more likely for the early seed times, we also show the probability of the variant not being used again after creation given by Eq.~\eqref{eq:q1_classic}. Lastly, Fig.~\ref{fig:class_simon_sims}(d) shows the average popularity of a variant seeded at time $\tau$ and observed at time $\Omega = 10^5$ along with the theoretical calculation given by Eq.~\eqref{eq:mean_classic}, demonstrating the power-law relationship between the two quantities. We conclude by commenting on the generally excellent agreement between simulations and the theoretical results offered by the branching process interpretation of Simon's classic model.

	\section{A generalized Simon's model}\label{sec:GSM}
	
	One of the enticing properties of Simon's original model is the simple manner in which previously used elements are chosen from when a reproduction event occurs. Specifically, the choice is made over all previously used elements such that the likelihood of choosing a certain variant is proportional to the number of times that variant has appeared prior to the choice, resulting in a preferential attachment form of dynamics. While such properties have resulted in an extensive array of literature in more recent times the concept of temporal memory~---~more formally known as the burstiness~\cite{Barabasi2005, Delvenne_2015, Karsai_2018}~---~underlying how one makes decisions when copying has become increasingly important from a modeling perspective.
	
	Accordingly, here we focus on such aspects by proposing a generalized Simon's model. In this model, when a new element from a reproduction event is choosing their variant type from the previously used elements the likelihood of copying from an element which was present~$t$ time units prior is proportional to a given function~$\phi(t)$. This probability may be viewed as a memory function through assuming the property~$\int_{0}^{\infty} \phi(w) \, dw = 1$. The quantity of interest  is, as before, the probability that a variant which first appeared at time~$\tau$ has been used~$n$ times by time~$\Omega$, denoted by~$q_n(\tau, \Omega)$. Incorporation of the temporal element of memory requires some further considerations when performing calculations for this quantity. Specifically, we start by considering the PGF~$H(\tau, a, \Omega; x)$ of the random variable describing the number of times that an element, seeded at time~$\tau$ due to a reproduction event from an element which had age~$a$ at the time (i.e., first appeared itself at time~$\tau - a$), has caused further elements to have the same variant type. Now, as in the classical case of Sec.~\ref{sec:classic}, we consider how this PGF changes over the time interval~$(\tau - \Delta t, \tau)$. If $\Delta t$ is sufficiently small such that at most one event occurs in said time interval, there are three possible events:
	\begin{enumerate}
		\item First, there may be a reproduction event w.p. $1-\mu$ where the new element chooses to copy the element of interest's variant by looking back to the time which they appeared w.p.~$\phi(a) \Delta t$. This results in a new element of age $a = 0$ with the same variant that may generate further progeny,  the total size of which would be added to the size of the previously seeded element and as such this event contributes $H(\tau, a, \Omega; x) H(\tau, 0, \Omega; x)$ to the PGF $H(\tau - \Delta t, a - \Delta t, \Omega; x)$.
		\item There may again be a reproduction event  w.p.~$1~-~\mu$ but in this instance the new element copies its variant from an alternative element w.p. $1 - \phi(a) \Delta t$ which does not affect the abundance resulting from the element of interest, thus the contribution to its PGF is simply $H(\tau, a, \Omega; x)$.
		\item Finally, there may be a mutation event rather than a reproduction w.p.~$\mu$ which again does not change the abundance size resulting from the element under consideration here thus contributes $H(\tau, a, \Omega; x)$ to the PGF.
	\end{enumerate}
	\begin{widetext}
		Taking these terms alongside their corresponding probabilities leads to the following difference equation describing the time evolution of the PGF
		\begin{equation}
			H(\tau - \Delta t, a - \Delta t, \Omega; x) = (1- \mu)\phi(a) H(\tau, a, \Omega; x)\left[H(\tau, 0, \Omega; x) - 1\right] \Delta t + H(\tau, a, \Omega; x).
		\end{equation}
		Taking, as in Sec.~\ref{sec:classic}, the case where $\Delta t \ll \tau$, and using a two-dimensional Taylor approximation gives the following partial differential equation
		\begin{equation}
			\frac{\partial H(\tau, a, \Omega; x)}{\partial \tau} + \frac{\partial H(\tau, a, \Omega; x)}{\partial a} = (1- \mu)\phi(a) H(\tau, a, \Omega; x)\left[1 - H(\tau, 0, \Omega; x)\right],
		\end{equation}
		which may be solved via the method of characteristics (we refer the interested reader to~\cite{OBrien2020} for a similar calculation) to obtain
		\begin{equation}
			\tilde{H}(t,a;x) = x\exp\left\{(1-\mu)\int_{0}^{t} \phi(u+a) \left[\tilde{H}(t-u,0; x) - 1\right] \mathrm{d}u \right\},
		\end{equation}
		where $\tilde{H}(\Omega-\tau, a; x) = H(\tau,a,\Omega;x)$, i.e., the functional form of this PGF now depends only upon the time interval $t = \Omega - \tau$ between seeding and observation.
		
		This PGF describes the distribution of abundance as a result of a new element's variant being chosen by copying from a previous element of age~$a$. We are generally interested in the first such seeding i.e., by an element with age $a=0$, to obtain information regarding the entire tree size, and as such we may drop the explicit use of $a$ and obtain  
		\begin{equation}
			\tilde{H}(t;x) = x \exp\left\{(1-\mu)\int_{0}^{t} \phi(u) \left[\tilde{H}(t-u; x) - 1\right] \mathrm{d}u \right\}.
			\label{eq:pgf_memory}
		\end{equation}
		We note the similarity between this distribution and Eq.~(8) of~\cite{OBrien2020}, indicating that the GSM may be viewed as a specific version of the self-exciting component within a Hawkes process~\cite{Hawkes_1971} for a given variant. As such, previous results regarding the predictability of future popularity size of said processes are readily applicable to abundance size in the case of GSM.This apparent equivalence is rather intriguing and demonstrates a deep underlying relationship between age-dependent branching processes and a range of random-copying continuous processes. We do note however that this equivalence arises from the branching process assumption, which describes the Hawkes process exactly, but in the case of Simon's model is an approximation as during each time step only one event should occur. This violates the independence assumption of the branching process, although in Appendix~\ref{app:errors} we demonstrate that the errors arising from this approximation in the classical model are minor.
	\end{widetext}
	
	\subsection{Analysis}
	Having obtained the functional form of the PGF through Eq.~\eqref{eq:pgf_memory} for the GSM we may now proceed to consider some possible analytical quantities obtainable regarding the process.
	\subsubsection{Mean Popularity}\label{subsubsec:GSM_mean}
	As in Sec.~\ref{subsubsec:classic_mean}, the mean abundance of a given variant after~$t$ time units have passed since it first appeared may be obtained directly from Eq.~\eqref{eq:pgf_memory} through the following calculation
	\begin{align}
		m(t) &= \left. \pd{\tilde{H}(t; x)}{x} \right\rvert_{x = 1} \nonumber \\
		&= 1 + (1-\mu)\int_{0}^{t}\phi(w) \, m(t - w) \, \mathrm{d}w,
		\label{eq:mean_memory}
	\end{align}
	which does not, in general, have an analytical solution. We may, however, conduct analysis in the Laplace space to determine properties of this quantity as follows
	\begin{equation}
		\hat{m}(s) = \frac{1}{s} + (1-\mu) \, \hat{\phi}(s) \, \hat{m}(s),
	\end{equation}
	where hatted quantities represent their Laplace transform $\hat{\phi}(s) = \int_{0}^{\infty} \phi(t) e^{-st} \, ds$. We  proceed to solve this equation for the Laplace-transformed average popularity  to obtain
	\begin{equation}
		\hat{m}(s) = \frac{1}{s\left[1 - (1-\mu) \hat{\phi}(s) \right]}.
		\label{eq:mhat}
	\end{equation}
	From this expression we may immediately calculate, by means of the final-value theorem, the average abundance in the infinite time limit as follows
	\begin{align}
		m(\infty) &= \lim_{t \to \infty} m(t) = \lim_{s \to 0} s \, \hat{m}(s) \nonumber \\
		&= \frac{1}{\mu}.
	\end{align}
	More generally (aside from specific distributions, e.g., the exponential, which may be handled analytically) the finite-time behavior of this quantity is obtained via  numerical inversion of the Laplace transform given by Eq.~\eqref{eq:mean_memory}. This is readily performed using the efficient Talbot algorithm~\cite{Talbot_1979,Abate_2006, Gleeson_2016}.
	
	One final consideration we describe here is the relationship between the mean memory-time $T = \int_{0}^{\infty} t \, \phi(t) \, dt$ and the corresponding large-time mean abundance. This is studied by considering the small-$s$ behavior of Eq.~\eqref{eq:mhat} in particular noting that 
	\begin{align}
		\hat{\phi}(s) &\approx \hat{\phi}(0) + s \, \frac{d \hat{\phi}(0)}{ds} \nonumber \\
		&= 1 - s \, T,
	\end{align}	
	where we have used the fact that~$\hat{\phi}(0)~=~\int_{0}^{\infty} \phi(t) \, dt~=~1$ and $\frac{d\hat{\phi}(0)}{ds}~=~\int_{0}^{\infty} -t \phi(t) \, dt~=~- T$. Using this approximation in Eq.~\eqref{eq:mhat} allows the mean abundance of a given variant~$t$ time units after first appearing to be written explicitly as 
	\begin{equation}
		m(t) \approx \frac{1}{\mu}\left[1 - e^{-\mu t/(1-\mu)T}\right], \qquad t \gg 1.
		\label{eq:mean_time_approx}
	\end{equation}
	Furthermore, in the case of an exponential memory kernel with mean time $\beta$ i.e., $\phi(t) = \frac{1}{\beta}e^{-t/\beta}$, we may exactly determine the mean abundance through inversion of Eq.~\eqref{eq:mhat} to obtain
	\begin{equation}
		m(t) = \frac{1}{\mu}\left[1 - (1-\mu)e^{-(\mu/\beta)t}\right].
		\label{eq:mean_exp}
	\end{equation}
	Indeed, this expression for the exponential memory kernel may be expanded to first order in the case where $t \ll \beta/\mu$ giving
	\begin{equation}
		m(t) \approx 1 + \frac{(1-\mu)}{\beta} t,
	\end{equation} 
	i.e., a linear function of time with an inverse dependence upon the mean memory-time of the distribution.
	
	\subsubsection{Infinite-age distribution}
	Analytically obtaining the probability distribution associated with the PGF given by Eq.~\eqref{eq:pgf_memory}, unlike that for the classical model, generally proves infeasible as a consequence of its convoluted nature. We may, however, conduct asymptotic analysis regarding the distribution in the case of those variants with large age~\cite{Gleeson2014}. Specifically we proceed to assume that $\tilde{H}(t;x)~\rightarrow~\tilde{H}_\infty(x)$,  independent of $t$, as $t \to \infty$ and make use of the fact that $\int_{0}^{\infty} \phi(w) \, \mathrm{d}w = 1$ in Eq.~\eqref{eq:pgf_memory} to obtain
	\begin{equation}
		\tilde{H}_\infty(x) = x \exp \left\{ (1-\mu) \left[\tilde{H}_\infty(x) - 1\right] \right\}.
		\label{eq:Hinfty}
	\end{equation}
	The large-$n$ behavior of the underlying distribution $q_n(\infty)$ may be studied by letting $x~=~1 - w$ and $\tilde{H}_\infty(x)~=~1~-~\gamma(w)$ before proceeding to make use of the small-$w$ and small-$\gamma$ asymptotics of Eq.~\eqref{eq:Hinfty}. Through this analysis (detailed in Appendix \ref{app:large-a}) it may be shown that the large-$n$ abundance distribution is described by a power-law with exponential cutoff
	\begin{equation}
		q_n(\infty) \sim A \, n^{-3/2} \, e^{-n/\kappa}, \quad \text{as $n \rightarrow \infty$},
		\label{eq:largeaqn_mu}
	\end{equation}
	where the prefactor $A$ is given by 
	\begin{equation}
		A = \sqrt{\frac{1-\mu}{2\pi}},
		\label{eq:largea_prefactor}
	\end{equation}
	and the cutoff scale $\kappa$ is 
	\begin{equation}
		\kappa = \frac{2(1-\mu)^2}{\mu^2}.
		\label{eq:largea_cutoff}
	\end{equation}
	We note that this heavy-tailed behavior is distinct from that found in Simon's original analysis, as in the original case Simon considered an abundance distribution across all variants within the system whereas we focus on a single variant. The present result however suggests that the incorporation of a decaying memory within the model results in a heavier tail describing the large-$n$ behavior of the abundance distribution even in the case of a single variant. A similar result has been shown in~\cite{Schaigorodsky2018} for a specific memory function (see Sec.~\ref{sec:special}) which also resulted in a power law with an exponent of -3/2 and exponential cutoff, the results presented here therefore demonstrate the generalization of this behavior to the case of an arbitrary memory kernel.

	\section{Simulations}\label{sec:simulations}
	
	We now proceed to consider different forms of the GSM via numerical simulations alongside the analytical estimates of their behavior obtained in the preceding section. In each case we generate sequences of elements similar to those in Sec.~\ref{subsec:classic_sim} but now, in the case of a replication event, the time from which to choose a target element for copying  is a random variable drawn from the specified memory distribution. Due to the generally infinite support of the GSM kernel there are times in which the copying time may occur prior to the start time of the system; we consider such cases as mutation events due to elements from prior to the system's start time not being of the same variant as that in which we are interested. 
	
\begin{figure}
		\centering
		\includegraphics[width = \columnwidth]{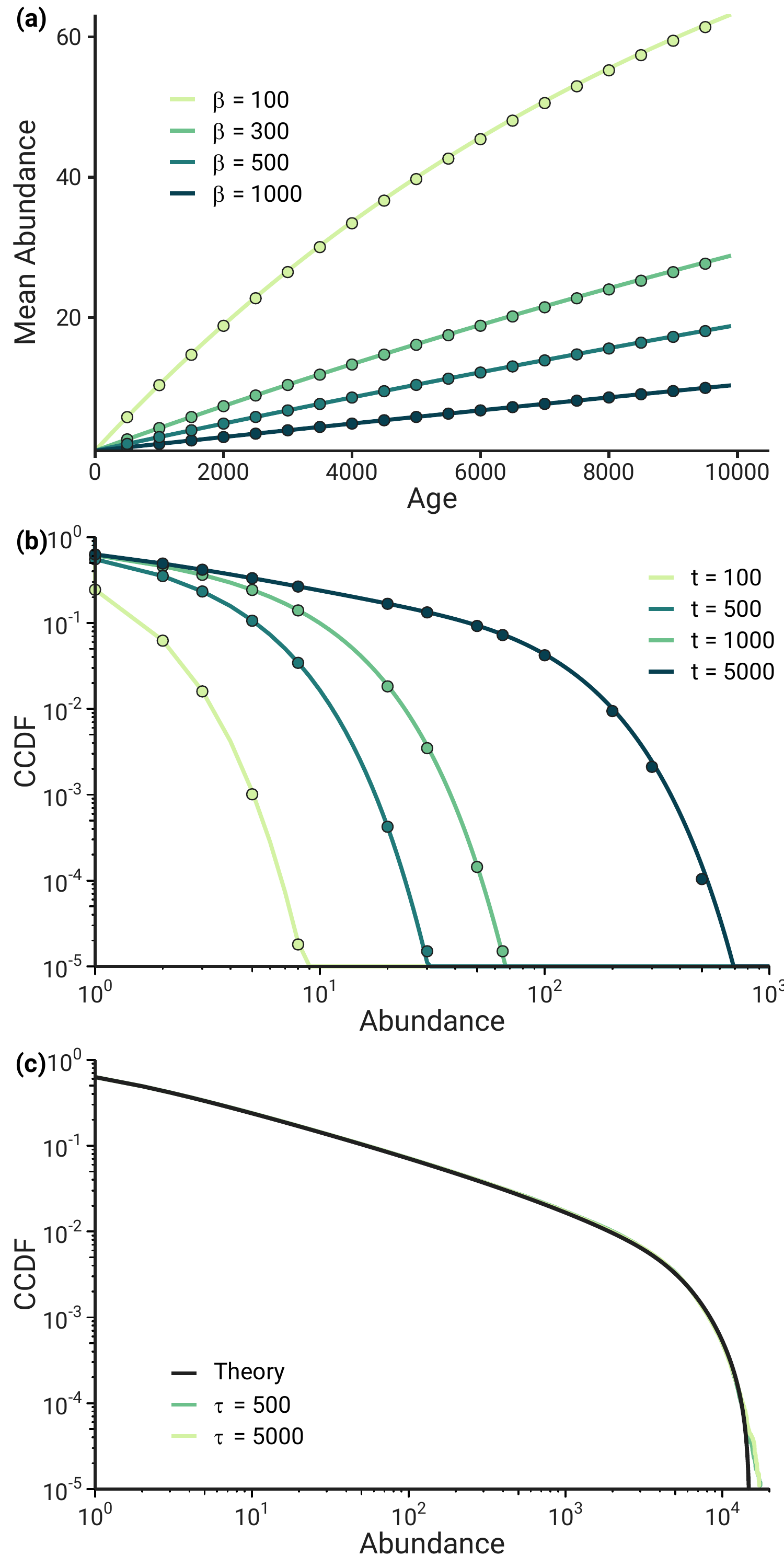}
		\caption{\textbf{Simulations and corresponding analytical results for the generalized Simon's model in the case of an exponential memory kernel. (a)}~Average abundance of a given variant as a function of its age obtained over $10^6$ realizations for various $\beta$ values and $\mu = 0.01$;  the dots represent simulated values and curves are the theoretical results. \textbf{(b)}~ CCDF of variant abundance at a number of different ages with $\beta = 300$, alongside theoretical obtained distributions from inversion of Eq.~\eqref{eq:pgf_memory}. \textbf{(c)}~CCDF of abundance for two variants which have been seeded at different times $(\tau = 500, 5000)$ but have been observed after the same amount of time has passed after seeding, resulting in identical distributions. Also note the power-law along with exponential cut-off behavior as predicted by Eq.~\eqref{eq:largeaqn_mu}.}
		\label{fig:GSM_exp}
	\end{figure}

	First, we simulate processes of length $10^4$ with an exponentially distributed memory kernel with mean time $\beta$ and innovation probability $\mu = 0.01$. The average number of times the variant has been copied by the time it has age~$t$, i.e., when there have been $t$ further elements created, is shown in Fig.~\ref{fig:GSM_exp}(a) for $\beta = 100, 300, 500,$ along with the theoretical prediction given by Eq.~\eqref{eq:mean_exp}. We point out the apparent inverse relationship between mean memory time and average abundance of a given variant, suggesting that shorter memory times favor the likelihood of a variant having larger popularity. We proceed to consider the distribution of abundances across these realizations in the case $\beta = 300$ at a number of different ages $t = 100, 500, 1000, 5000$ in Fig.~\ref{fig:GSM_exp}(b), where the complementary cumulative distribution function is shown along with the corresponding theoretical estimates obtained from inversion of Eq.~\eqref{eq:pgf_memory}. Lastly we consider, as in Fig.~\ref{fig:class_simon_sims}(b), two variants seeded at different times $\tau = 500, 5000$ and observe the abundance of the variant at $\Omega = \tau + 50000$ i.e., when they both have the same age. Now, unlike in the classical Simon's model, the distributions are equivalent such that there is no longer a `early-mover' advantage for variants.
	
\begin{figure}
		\centering
		\includegraphics[width = \columnwidth]{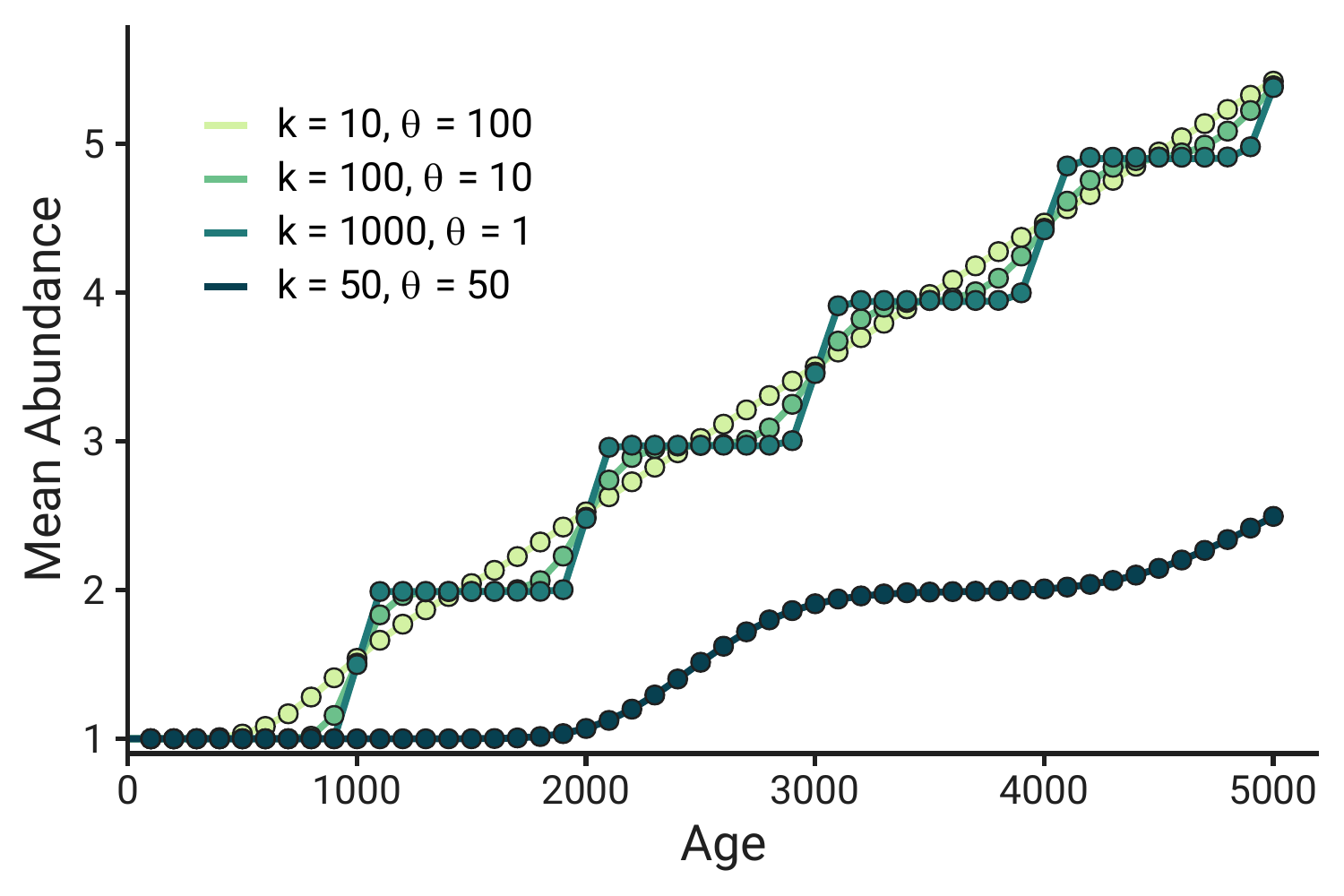}
		\caption{\textbf{Simulations and corresponding analytical results for the average abundance of the generalized Simon's model in the case of a Gamma distributed memory kernel.} A number of parameter combinations for the distribution are used. We see how the general trend of the average abundance is represented through the mean $k \theta$ of the Gamma distribution while the temporal dynamics of the abundances fluctuate based upon its shape. In each case the average is computed over $10^6$ realizations with $\mu = 0.01$.}
		\label{fig:GSM_gamma}
	\end{figure}

	We now proceed to consider an alternative memory kernel to demonstrate the robustness of our GSM formalism, specifically we consider a gamma-distributed memory given by
	\begin{equation}
		\phi(t) = \frac{t^{k-1}e^{-t/\theta}}{\theta^k\Gamma(k)},
	\end{equation}
	where $k$ is called the shape parameter and $\theta$ the scale parameter. This distribution offers a wide-range of behaviors, the effect of which upon the GSM may be observed. In particular, the average memory-time is given by $T = k\theta$ while the structure of the distribution can vary widely from exponential (when $k=1$) to more
	symmetrically distributed about the mean when $k > \theta$. Figure~\ref{fig:GSM_gamma} again shows the average abundance of a variant simulated with this process over $10^6$ realizations. Two features are immediately noted, firstly we see that the general trend of the abundance is dependent upon the mean memory-time whereby the cases where $k = 10,100,1000$ and $\theta = 100,10,1$ all have a equivalent general trend while the process with larger mean memory-time ($k = \theta = 50$) demonstrates greater average abundance growth. The second interesting aspect  here is the effect 
	of the shape of the memory distribution
	upon the dynamics of the average abundance. In particular, the more skewed scenario ($k = 10, \theta = 100$) results in linear-like growth as observed in the exponentially distributed case. On the other hand, when the memory distribution is more symmetric the likelihood of a new element replicating the variant of interest is most likely at time units close the average memory times, resulting in the almost step-wise behavior of the averages. Importantly, in this case there is no closed-form solution for the theoretical expression of the expected abundance and  the results are obtained via numerical inversion of Eq.~\eqref{eq:mhat}.
	
	\section{Special cases}\label{sec:special}
	
	In the preceding section we proposed the incorporation of a general memory kernel within a Simon's model framework. Prior literature has considered specific kernels, including Refs.~\cite{Cattuto2006,Cattuto2007} where a hyperbolic memory was studied, and Ref.~\cite{Schaigorodsky2018} which, like the original Simon's model,  utilizes a uniform memory that is, however, bounded such that elements are not considered after a certain amount of time has passed.  Here we demonstrate how our proposed framework can describe these generalizations (particularly in the case of the latter) in continuous time while also allowing the determination of additional information regarding the statistical properties of the abundance distribution.
	
	\subsection{Schaigorodsky et al.}
	
	In~\cite{Schaigorodsky2018} Schaigorodsky et al.~propose an extension to Simon's model which they coin bounded memory preferential growth (BMPG) whereby their choice of memory kernel is given by a step function
	\begin{equation}
		\phi(t) =
		\left\{
		\begin{array}{cc}
			\frac{1}{\kappa} & \mathrm{if\ } t \le \kappa, \\
			0 & \mathrm{if\ } t > \kappa. \\
		\end{array} 
		\right.
		\label{eq:BMPG_memory}
	\end{equation}
	This implies that the variant of a new element is uniformly chosen from the previous $\kappa$ elements (with $\Delta t = 1$). We may proceed, as before, to consider the average popularity of a variant with age $t$ by calculating the Laplace transform of Eq.~\eqref{eq:BMPG_memory} given by
	\begin{equation}
		\hat{\phi}(s) = \frac{1}{\kappa s}\left[1 - e^{-\kappa s}\right] ,
	\end{equation}
	before substituting in Eq.~\eqref{eq:mhat} to obtain
	\begin{equation}
		\hat{m}(s) = \frac{1}{s - \frac{(1-\mu)}{\kappa}\left[1 - e^{-\kappa s}\right]},
		\label{eq:BMPG_mhat}
	\end{equation}
	which gives the Laplace transform of the mean abundance size at a  time $t$ after the variant first appears in the BMPG model. While this quantity itself does not have an analytical inverse transform, it is still possible to invert it numerically as described in Sec.~\ref{subsubsec:GSM_mean}. Furthermore, we may perform approximate analysis regarding the mean abundance, as per Eq.~\eqref{eq:mean_time_approx}, which is a valid approximation here in the regime where $t \gg \kappa$. This calculation depends on the mean-memory time which in the case of BMPG is given by $T = \kappa/2$ and the corresponding expected abundance for a variant with age $t$ is given by
	\begin{equation}
		m(t) \approx \frac{1}{\mu}\left[1 - e^{\frac{-2\mu}{(1-\mu)\kappa}t}\right].
	\end{equation}
	If we then consider the Taylor series of this function about $t = 0$, particularly focusing on the linear term, which is valid when $t \ll \frac{(1-\mu)}{2\mu}\kappa$, we obtain
	\begin{align}
		m(t) &\approx \frac{1}{\mu}\left[1 - \left(1 - \frac{2\mu}{(1-\mu)\kappa}t\right)\right] \nonumber \\
		&\approx \frac{2}{(1-\mu)\kappa}t, \qquad \kappa \ll t \ll \frac{(1-\mu)}{2\mu}\kappa.
		\label{eq:BMPG_earlyt_mean}
	\end{align}
	
	Furthermore, we readily obtain the entire distribution of abundances for variants of any age by substituting Eq.~\eqref{eq:BMPG_memory} into Eq.~\eqref{eq:pgf_memory} and inverting numerically as before, from which moments of the distribution may also be calculated.
	
	\begin{figure}[t!]
		\centering
		\includegraphics[width = \columnwidth]{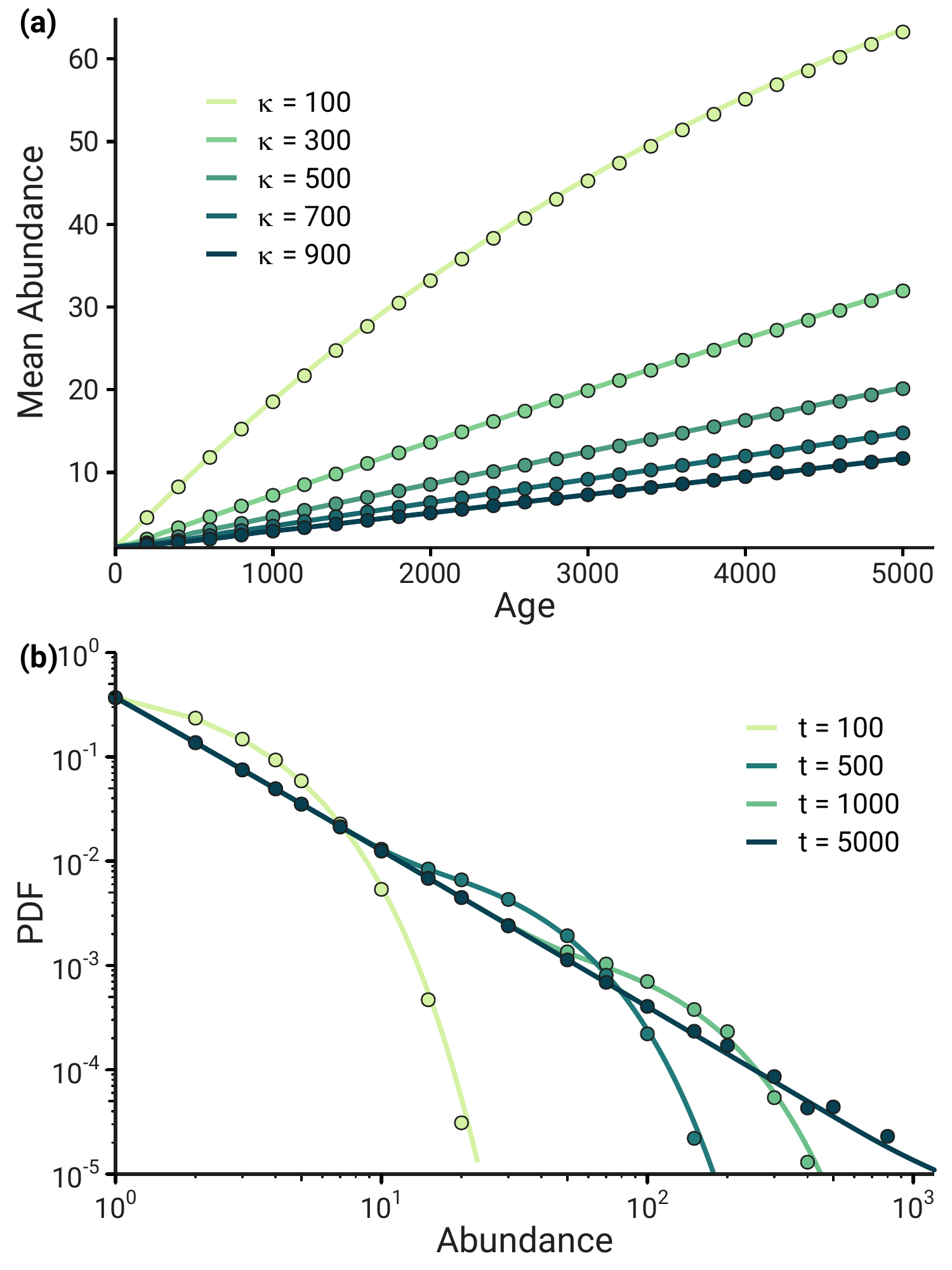}
		\caption{\textbf{Analysis of the bounded memory preferential growth (BMPG) model alongside analytical estimates from the branching process description. (a)}~Average abundance of a given variant as a function of its age obtained over $10^6$ realizations for a number of $\kappa$ values representing the length of the memory kernel. Here $\mu = \kappa^{-1}$ and the theoretical curves are from Eq.~\eqref{eq:BMPG_mhat}. \textbf{(b)}~PDF of variant abundance at a number of different ages in the case  $\kappa = 100$, with theoretical  distributions obtained from inversion of Eq.~\eqref{eq:pgf_memory}.}
		\label{fig:BMPG_sims}
	\end{figure}
	
	In order to validate the applicability of our GSM to this scenario we proceed to conduct numerical simulations of the process, with results  shown in Fig.~\ref{fig:BMPG_sims}. Each simulation is averaged over $10^6$ realizations with the dots representing simulated values and the curves their theoretical equivalents. Figure~\ref{fig:BMPG_sims}(a) shows the average abundance of a variant for various $\kappa$ values where, as per \cite{Schaigorodsky2018}, in each case the innovation probability $\mu$ is the reciprocal of the memory length $\kappa$. The theoretical results are determined via numerical inversion of Eq.~\eqref{eq:BMPG_mhat} and generally offer excellent agreement with simulation. We also observe that in this case the linear approximation offered by Eq.\eqref{eq:BMPG_earlyt_mean} holds within the time regime $\kappa \ll t \ll \kappa^2/2$, which supports the linear behavior observed in the mean abundance and furthermore gives an estimate for the slope of these functions to be approximately $2/\kappa$.  The corresponding probability distribution functions of abundance size  at a number of different ages are shown in Fig.~\ref{fig:BMPG_sims}(b) with $\kappa = 100$. We also observe the `peloton dynamics' reported in~\cite{Hebert-Dufresne2012} whereby there are bulges within the distribution as it converges towards its large-time limit.
	
	\subsection{Cattuto et al.}
	
	A further study which analyses a memory-dependent Simon's model may be found in the works~\cite{Cattuto2006, Cattuto2007} by Cattuto et al. in which the memory kernel is defined as
	\begin{equation}
		\phi(t) = \frac{C}{\gamma + t},
	\end{equation}
	where $\gamma$ represents a timescale characteristic  to the system under consideration and $C$ is simply a normalization factor determined by the condition $\int_0^{t_\text{max}} \phi(w) \, dw=1$, 
	where $t_\text{max}$ is the current total time of the process, i.e., the memory distribution is varying at each step. This specific time-dependent memory kernel is beyond the scope of our proposed framework in that we assume a fixed kernel defined such that $\int_0^\infty \phi(w) \, dw = 1$. We may, however, consider a similar constant kernel for which the support is bounded by some upper time limit $t^*$ whereby $C$ is determined to ensure $\int_0^{t^\star} \phi(w) \, dw = 1$. Numerical simulations (omitted from the present article) of the GSM with this memory kernel, like those studied above, offer perfect agreement with theoretical estimates of the abundance size distribution obtained from inversion of Eq.~\eqref{eq:pgf_memory}. Exact agreement with the original model proposed in~\cite{Cattuto2006, Cattuto2007} however would require the incorporation of a time-varying memory function within our framework which is beyond the scope of the present article.
	
	\section{Conclusions}\label{sec:conclusions}
	
	In this paper we have proposed a branching process framework to describe Simon's renowned random-copying model~\cite{Simon1955} and generalizations thereof. This mapping is determined by identifying the replication events of the original model as equivalent birth events in the branching process. Through this description we obtain a detailed understanding of the temporal evolution of a new variant's abundance via a knowledge of its probability distribution under the branching process assumption. Using this analysis we reproduce classical results while also demonstrating the potential to determine further statistical properties. These findings are validated via extensive numerical simulations which ratify this approach in describing the model. Interestingly, in spite of the model's discrete nature, we find that the continuous time description suggested by our framework is extremely useful in further understanding properties of the original model.
	
	Motivated by the branching process framework accurately describing Simon's original model (see Appendix~\ref{app:errors} for a discussion on the accuracy of this interpretation) we proceeded to extend beyond the classical uniform random-copying probability to any arbitrary age-dependent memory kernel resulting in our GSM framework. Such considerations have previously been studied in the case of specific kernels~\cite{Cattuto2006,Cattuto2007,Schaigorodsky2018, Bertoin_2019} and we now extend this to the general scenario. Using these ideas we demonstrate the potential for rich behaviors in such models that are in strong agreement with numerical simulations. Indeed, as per the classical model, it is possible to determine statistical properties of the process via analytical (or semi-analytical) calculation rather than through extensive Monte Carlo simulations. Lastly, we highlight the fact that power-law distributions of abundance size, as originally highlighted in Simon's model, occur for a different reason in the GSM for a variant with large age, albeit now with a smaller exponent ($3/2$) and an exponential tail.
	
	Given the ubiquitous use of Simon's model as a basis for a range of different frameworks, including growing networks~\cite{Barabasi1999, Hebert-Dufresne2012} and in describing the abundance distributions in numerous empirical systems~\cite{Zanette2001, Bornholdt2001, Maillart2008}, we believe that the theory introduced in this article offers the potential for much further exciting research.
	An important example is  the incorporation of memory effects~\cite{Gleeson_2016, Karsai_2018} within models for dynamical systems describing the behaviors of human activity. Lastly, while Simon's model is arguably the most notable neutral model~\cite{Leroi2020}, there are many other similar models which we anticipate to benefit from our framework in further understanding the temporal dynamics of variant abundance size. 
	
	\begin{acknowledgements}
		Helpful comments from two anonymous reviewers are gratefully acknowledged.~This work was supported by Science Foundation Ireland Grant No.~16/IA/4470, No.~16/RC/3918, and No.~12/RC/2289P2 (J.D.O'B. and J.P.G.).  We acknowledge the DJEI/DES/SFI/HEA Irish Centre for High-End Computing (ICHEC) for the provision of computational facilities and support.
	\end{acknowledgements}
	
	\bibliography{out}
	
	\appendix
	
	\appendix
	\numberwithin{equation}{section}
	\numberwithin{figure}{section}
	
	\section{Alternative derivations of Eq.~\ref{eq:Hclassic}}
	We shall now consider alternative approaches to the PGF, given by Eq.~\eqref{eq:Hclassic} in the main text, from which the probability distribution associated with the classical Simon's model may be derived. First we define $q_n(t)$ to be the probability of a specific variant having abundance~$n$ at time~$t$. We now consider the different ways in which the variant could have appeared~$n$ times by time~$t$:
	\begin{enumerate}
		\item The variant could have appeared $n-1$ times by time $t$ and then be chosen again as a result of a new element replicating from one of the previous elements, with probability  $(1-\mu)\left(\frac{n-1}{t}\right)$.
		\item The variant could have instead appeared $n$ times by time $t$ and a new element again chooses to replicate, but this time chose an alternative variant with probability $(1-\mu)\left(1 - \frac{n}{t}\right)$.
		\item Finally, the variant could have appeared $n$ times by time $t$ and the new element was a mutation with probability $\mu$, thus not creating another copy of the variant in question.
	\end{enumerate}
	Combining these possibilities gives us the following master equation:
	\begin{align}
		q_n(t+1) &= (1-\mu)\left(\frac{n-1}{t}\right) q_{n-1}(t) + \nonumber \\
		& \qquad (1-\mu)\left(1 - \frac{n}{t}\right) q_n(t) + \mu q_n(t), \nonumber \\
		&= q_n(t) + \frac{1-\mu}{t+1}\left[(n-1)q_{n-1}(t) - nq_n(t)\right].
		\label{master}
	\end{align}
	With this equation at hand, we now proceed to consider two mathematical approaches from which we may obtain the corresponding PGF.
	
	\subsection{Generating function approach}
	Defining $G(t;x) = \sum_{n}^{}q_n(t)x^n$ to be the PGF of the random variable $q_n(t)$, we then multiply Eq.~(\ref{master}) by $x^n$ and sum over all $n$ to obtain
	\begin{equation}
		G(t+1; x) = G(t;x) + \frac{1-\mu}{t}(x^2 - x)\pd[]{G(t;x)}{x},
	\end{equation}
	and considering the large-time case such that the change in one time unit is relatively small we obtain
	\begin{equation}
		\pd[]{G(t;x)}{t} = \frac{1-\mu}{t}\left(x^2 - x\right)\pd[]{G(t;x)}{x}.
		\label{pde_master}
	\end{equation}
	We now also note that the initial condition governing this first order homogeneous partial differential equation (PDE) is dependent on when the variant first appeared in the sequence, which we shall call time $\tau$, such that $G(\tau; x) = x$. To solve this PDE we use the method of characteristics noting that the two equations governing the shape of the characteristic curve are given by:
	\begin{equation}
		\frac{dt}{dr} = \frac{t}{1-\mu}, \qquad \frac{dx}{dr} = x(1-x),
	\end{equation}
	which are readily solved to give
	\begin{equation}
		r = (1-\mu)\log(t), \qquad r = \log\left(\frac{x}{1-x}\right) + C,
	\end{equation}
	where $C$ is the constant of integration. We now note that due to the PDE being homogeneous, $G(t;x)$ is a constant along the characteristic curve and we can deduce the general solution to be given by
	\begin{equation}
		G(t;x) = F(C) = F\left[(1-\mu)\log(t) - \log\left(\frac{x}{1-x}\right)\right].
	\end{equation} 
	As we are aware of the initial condition of the differential equation, i.e., $G(\tau;x) = x$, we may determine the functional form of $F$ which allows us to conclude:
	\begin{equation}
		G(t;x) = \frac{x}{x\left[1 - \left(\frac{\tau}{t}\right)^{-(1-\mu)}\right] + \left(\frac{\tau}{t}\right)^{-(1-\mu)}},
	\end{equation}
	in agreement with the PGF given by Eq.~\eqref{eq:Hclassic} obtained via the branching process analysis.
	
	\subsection{Inductive approach}
	Another possible approach to obtaining this distribution is to take Eq.~(\ref{master}) and consider the small time change without directly transforming to a PGF in order to obtain
	\begin{align}
		q_n'(t) + n\left(\frac{1-\mu}{t}\right)q_n(t) &= (n-1)\left(\frac{1-\mu}{t}\right)q_{n-1}(t),
	\end{align}
	which implies
	\begin{align}
		\frac{d}{d t}\left[t^{n(1-\mu)}q_n(t)\right] &= t^{n(1-\mu)}\left(\frac{1-\mu}{t}\right)(n-1)q_{n-1}(t).
		\label{ode}
	\end{align}
	
	Again we note the dependence on the seed time $\tau$, as $q_1(\tau) = 1$ and $q_0(t) = 1, \  \forall t < \tau$. Solving this equation we find the probability of a variant which first appeared at time $\tau$ having popularity~$n$ at time~$t$ to be given by
	\begin{equation}
		q_n(t) = \left(\frac{\tau}{t}\right)^{1-\mu}\left[1 - \left(\frac{\tau}{t}\right)^{1-\mu}\right]^{n-1},
	\end{equation}  
	as per Eq.~\eqref{eq:qn_classic} determined via the branching process description.
	
	\section{Errors from the branching process approximation}\label{app:errors}
	
	Throughout Sec.~\ref{sec:classic} of the main text we describe the classical Simon's model via a branching process approximation. It is important to note that this approach is, as stated, an approximation and in particular we consider the regime in which the time-step is smaller than the observation time, i.e., $\Delta t \ll \Omega$, enabling a continuous time analysis to be conducted. Simon's model is, however, a discrete process by nature. In spite of this we demonstrate, via numerical simulations, that the approach appears to work extremely well in terms of capturing the behavior of the abundance of a given variant. Here we consider the branching process approximation but in a discrete-time setting describing variants which appear near the observation time, for which the exact probabilities in both the BP setting and the model itself may be calculated and compared to determine the error associated with the approximation.
	
	We define $G_\tau(\Omega) = G(\tau, \Omega; x)$ to be the PGF for the probability mass function describing the abundance of a variant which appeared at time-step~$\tau$ observed at time~$\Omega$. Using similar arguments to those found in the main text we obtain the following difference equation to describe the dynamics of the process
	\begin{equation}
		G_\tau(\Omega) = G_{\tau+1}(\Omega) + \frac{(1-\mu)}{\tau}\left\{\left[G_{\tau+1}(\Omega)\right]^2 - G_{\tau+1}(\Omega)\right\},
		\label{eq:discrete_pgf}
	\end{equation}	
	where the squared term arises from the fact that a successful copying event can be construed as two separate trees beginning due to each element having the same probability of being copied in the classical model. Furthermore, we are aware of the final condition of this equation namely $G_\Omega(\Omega) = x$, i.e., a variant seeded at the observation time must have only appeared once. This equation, while due to its non-linear nature being analytically troublesome, may be comfortably studied symbolically.
	
	The approach taken above may seem extremely close to that underlying Simon's original model however the branching process description is not exact due to the fact that each tree created as a consequence of a successful copying event must be treated independently of one another from the branching process assumption. In the model itself, on the other hand, these trees are not independent as only one event may happen in a time step, i.e., if one element is successfully copied the other can not be.
	
	In order to study the effect of this assumption in the branching process model we consider a variant which appeared at time~$\tau = \Omega-2$ and we are concerned with the probability mass function of its abundance at time $\Omega$, i.e., after three potential copying events. Exact calculation of the quantities in the classical model,~$q_n^{\text{exact}}(\Omega - 2, \Omega)$ requires one to consider the possible sample paths for a variant to be copied $n-1$ times, for example we now demonstrate how one may calculate $q_2^{\text{exact}}(\Omega - 2, \Omega)$: 
	\begin{widetext}
		\begin{equation}
			q_2^{\text{exact}}(\Omega - 2, \Omega) = \left\{\underbrace{\frac{1-\mu}{\Omega-2}}_{\btext{Successful copy \\ at time $\Omega-1$}} \times \underbrace{\left[\mu + (1-\mu)\left(\frac{\Omega-3}{\Omega-1}\right)\right]}_{\btext{Unsuccessful copy \\ at time $\Omega$}}\right\} + \left\{\underbrace{\left[\mu + (1-\mu)\left(\frac{\Omega-3}{\Omega-2}\right)\right]}_{\btext{Unsuccessful copy \\ at time $\Omega-1$}} \times \underbrace{\frac{1-\mu}{\Omega-1}}_{\btext{Successful copy \\ at time $\Omega$}}\right\}.
		\end{equation}
		Clearly, this approach itself becomes computationally expensive as we move the seeding event further backwards in time. Similarly, we iterate Eq.~\eqref{eq:discrete_pgf} to obtain the equivalent probabilities for the branching process interpretation,~$q_n^\text{BP}(\Omega-2,\Omega)$, which are the coefficients of the resulting polynomial in $x$. Table.~\ref{tab:discrete_errors} demonstrates the probability values obtained from both approaches alongside the error resulting from the branching process approximation. Importantly, the error scales inversely with the observation time to the power of the number of time steps~---~three in the above example~---~which, in the case of large observation time, explains the good agreement observed between the branching process approximation and numerical simulations.
		
		\begin{table*}[t]
			\caption{\label{tab:discrete_errors}\textbf{Analysis of errors associated with the branching process approximation of Simon's classical model.} The distribution of variant abundance from a mutation event at time $\Omega - 2$ observed at time $\Omega$ is determined. The probabilities are calculated via both the branching process model from Eq.~\ref{eq:discrete_pgf} and the exact values via the sample paths of the true process, alongside the error calculated as the difference between the two.}
			\begin{tabular*}{\textwidth}{c @{\extracolsep{\fill}}ccc}
				\toprule
				\textbf{Abundance} $\mathbf{n}$ & $\mathbf{q_n^\text{\textbf{exact}}(\Omega-2,\Omega)}$ & $\mathbf{q_n^\text{\textbf{BP}}(\Omega-2,\Omega)}$ & \textbf{Error} \\
				\cmidrule{1-1}\cmidrule{2-2}\cmidrule{3-3}\cmidrule{4-4}
				1 & $\dfrac{(\Omega-3+\mu)(\Omega-2+\mu)}{(\Omega-2)(\Omega-1)}$ &$\dfrac{(\Omega-3+\mu)(\Omega-2+\mu)}{(\Omega-2)(\Omega-1)}$ & 0 \\
				2 & $\dfrac{(1-\mu)(2\Omega-6+3\mu)}{(\Omega-2)(\Omega-1)}$ & $\dfrac{(1-\mu)(\mu^2 + 2\Omega^2 + 3\mu\Omega-5\mu-8\Omega+7)}{(\Omega-2)(\Omega-1)^2}$ & $\dfrac{-(1-\mu)^3}{(\Omega-2)(\Omega-1)^2}$\\
				3 &  $\dfrac{2(1-\mu)^2}{(\Omega-2)(\Omega-1)}$ & $\dfrac{2(1-\mu)^2(\Omega + \mu - 2)}{(\Omega-2)(\Omega-1)^2}$ & $\dfrac{2(1-\mu)^3}{(\Omega-2)(\Omega-1)^2}$ \\
				4 & 0 & $\dfrac{(1-\mu)^3}{(\Omega-2)(\Omega-1)^2}$ & $\dfrac{-(1-\mu)^3}{(\Omega-2)(\Omega-1)^2}$ \\
				\bottomrule
			\end{tabular*}
		\end{table*}
		
	\end{widetext}
	
	\section{Derivation of large popularity behavior}\label{app:large-a}
	The large $n$ behavior of $q_n(a)$ may be obtained in the $a\to\infty$ limit by asymptotic analysis of Eq.~(\ref{eq:Hinfty}) around $x = 1$, via a similar approach as taken in Sec.~S3 of~\cite{Gleeson2014}.  Specifically, for $x$ near $1$ we let $w = 1-x$ and suppose $\tilde{H}_\infty(x) \sim 1 - \psi(w)$ as $w\to 0$. Now Eq.~(\ref{eq:Hinfty}) may be expressed as
	\begin{equation}
		1-\psi = (1-w)\exp\left\{-(1-\mu)\,\psi\right\},
	\end{equation}
	
\begin{figure}[t]
		\centering
		\begin{tikzpicture}
			\def\gap{0.2}
			\def\bigradius{3}
			\def\littleradius{0.5}
			\def\dist{1}
			
			\draw [help lines,->] (-1.2*\bigradius, 0) -- (1.2*\bigradius,0);
			\draw [help lines,->] (0, -1.2*\bigradius) -- (0, 1.2*\bigradius);
			\draw[blue, line width=1pt, decoration={ markings,
				mark=at position 0.2455 with {\arrow[line width=1.2pt]{>}},
				mark=at position 0.79 with {\arrow[line width=1.2pt]{>}},
				mark=at position 0.9 with {\arrow[line width=1.2pt]{>}},
				mark=at position 0.96 with {\arrow[line width=1.2pt]{>}}},
			postaction={decorate}]
			let
			\n1 = {asin(\gap/2/\bigradius)},
			\n2 = {asin(\gap/2/(\littleradius+\dist))},
			\n3 = {sqrt(\dist^2 + \littleradius^2 -2*\dist*\littleradius*cos(180 - asin(\gap/2/\littleradius)))},
			\n4 = {asin(\gap/2/\littleradius)}
			in (\n1:\bigradius) arc (\n1:360-\n1:\bigradius)
			-- ({-asin(\gap/2/\n3)}:\n3) arc (-\n4:-360+\n4:\littleradius)
			-- cycle
			;
			
			\node at (1.25*\bigradius,-0.25){$\Re(x)$};
			\node at (-0.4,1.25*\bigradius) {$\Im(x)$};
			\node at (0.35,0.43) {$C_{\varepsilon}$};
			\node at (-1.8,2.8) {$C_{R}$};
			\node at (2.6*\dist,0.29) {$l_1$};
			\node at (2.6*\dist,-0.32) {$l_2$};
			\node[label={270:{$\alpha$}},circle,fill,inner sep=2pt] at (\dist,0) {};
		\end{tikzpicture}
		\caption{The contour C in the complex x-plane for the PGF inversion integral Eq.(\ref{contour_int}). A branch cut extends from $\alpha$ to $\infty$.}
		\label{contour}
	\end{figure}
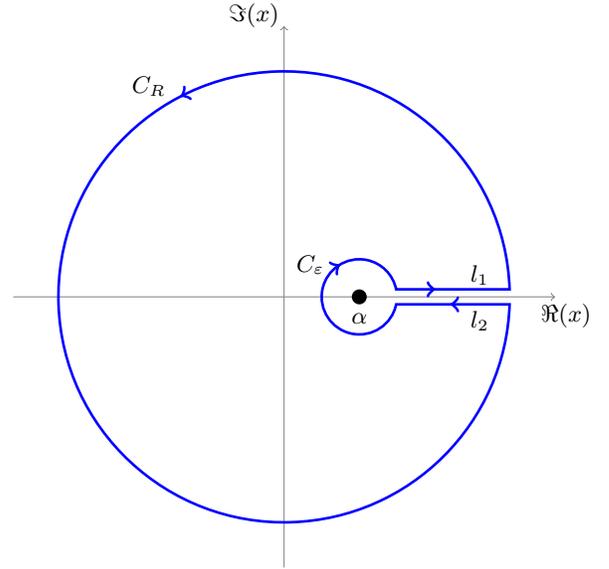

	Now let us consider the case where $0<\mu\ll1$, again taking $\tilde{H}_\infty(1-w) = 1 - \psi(w)$ and Taylor expanding Eq.~\eqref{eq:Hinfty} about $w = 0$ and keeping terms of order $w$, $\psi$ and $\psi^2$ but ignoring terms of order $\psi w$ we obtain
	\begin{equation}
		\frac{1}{2}(1-\mu)^2\psi^2 + \mu\psi - w = 0,
	\end{equation}
	which has solution
	\begin{equation}
		\psi = \frac{-\mu + \sqrt{\mu^2+2w(1-\mu)^2}}{(1-\mu)^2},
	\end{equation}
	where we have taken the positive root as the PGF should be bounded above by one. Note that a branch point $\alpha$ exists in the complex $x$ plane at $x = 1 + \frac{\mu^2}{2(1-\mu)^3}$. Now, using the Cauchy theorem we may express the discrete probabilities in terms of the underlying generating function as
	\begin{equation}
		q_n = \frac{1}{2\pi i}\oint_C \frac{\tilde{H}_\infty(x)}{x^{n+1}} \, \mathrm{d} x,
		\label{contour_int}
	\end{equation}
	where $C$ is a closed contour in the complex plane which does not enclose any poles of $\tilde{H}$. We deform $C$ such that it does not enclose the branch point $\alpha$ into the contour $C_R \cup l_2 \cup C_\varepsilon \cup l_1$ as seen in Fig.~\ref{contour}.
	It can be easily shown that the contributions to the integral from the circular contours $C_R$ and $C_\varepsilon$ limit to zero as $R\to\infty$ and~$\varepsilon~\to~0$~\cite{Keener2018}. Also, the contribution from the analytical part of $\tilde{H}$ will cancel due to the opposite directions of the line integrals. Therefore as $w \to 0$ we have
	\begin{align}
		q_n &= \frac{-1}{2\pi i} \int_{l_1 \cup l_2}^{} \frac{ \sqrt{\mu^2+2(1-x)(1-\mu)^2}}{(1-\mu)^2 \,x^{n+1}} \ \mathrm{d} x \\
		&= \frac{1}{2\pi i} \int_{l_1 \cup l_2}^{} \frac{\sqrt{\mu^2+2w(1-\mu)^2}}{(1-\mu)^2 \,(1-w)^{n+1}} \ \mathrm{d} w, \nonumber
	\end{align}
	now letting $w = 1 - e^\rho \approx -\rho$ as $\rho \to 0$ we obtain
	\begin{equation*}
		q_n = \frac{-1}{2\pi i} \int_{l_1 \cup l_2}^{}\frac{\sqrt{\mu^2-2\rho(1-\mu)^2}}{(1-\mu)^2} \ e^{-n\rho} \ \mathrm{d} \rho,
	\end{equation*}
	and making the substitution $\nu = \mu^2 - 2(1-\mu)^2\rho$ gives
	\begin{equation*}
		q_n = \frac{1}{4\pi i(1-\mu)^4}  \int_{l_1 \cup l_2}^{} \sqrt{\nu} \, e^{-n\left(\frac{\mu^2-\nu}{2(1-\mu)^2}\right)} \ \mathrm{d} \nu.
	\end{equation*}
	We now make the substitution $\nu = re^{i\theta}$ where along $l_1$ $\theta = \pi$, $\nu = -r$ and $\sqrt{\nu} = \sqrt{r}\,e^{\frac{i\pi}{2}}$ and along $l_2$ $\theta = -\pi$, $\nu = -r$ and $\sqrt{\nu} = \sqrt{r}\,e^{\frac{-i\pi}{2}}$, taking the limit as $R \to \infty$ we now consider the limits on the integral. In the $x$-plane $l_1$ varies from $1 + \frac{\mu^2}{2(1-\mu)^2}$ to $\infty$ and as such $w$ varies from $-\frac{\mu^2}{2(1-\mu)^2}$ to $-\infty$, hence $\rho$ takes values between $\frac{\mu^2}{2(1-\mu)^2}$ and $\infty$, and finally $\nu$ varies between $0$ to $-\infty$. A similar line of thought allows one to see that along $l_2$, $\nu$ varies between $-\infty$ to $0$. Thus our integral may now be expressed as
	\begin{equation}
		q_n = \frac{1}{2\pi(1-\mu)^4}e^{\frac{-n\mu^2}{2(1-\mu)^2}}\left[\frac{e^{\frac{i\pi}{2}} - e^{\frac{-i\pi}{2}}}{2i}\right] \int_{0}^{\infty}r^{1/2} e^{\left(\frac{-n}{2(1-\mu)^2}\right)r} \ \mathrm{d} r,
	\end{equation}
	and lastly letting $t = \frac{n}{2(1-\mu)^2}r$ we obtain
	\begin{equation}
		q_n =  \frac{\sqrt{2}}{\pi}(1-\mu)^{1/2}n^{-3/2}e^{\frac{-n\mu^2}{2(1-\mu)^2}}  \int_{0}^{\infty}t^{1/2} e^{-t} \ \mathrm{d} t,
	\end{equation}
	where the integral is now well-known with solution 
	\begin{equation}
		q_n = \frac{(1-\mu)^{1/2}}{\sqrt{2\pi}}n^{-\frac{3}{2}} e^{\frac{-n\mu^2}{2(1-\mu)^2}}
	\end{equation}
	
\end{document}